\documentclass[twocolumn,aps,pra]{revtex4}
\usepackage{amssymb}
\usepackage{graphicx}
\usepackage{dcolumn}
\usepackage{bm}
\usepackage{amsmath}
\usepackage[colorlinks,linkcolor=red,citecolor=blue]{hyperref}
\usepackage[linesnumbered,ruled]{algorithm2e}
\begin{document}

\title{The Bayesian Committee Approach for Computational Physics Problems}

\author{Li Chen$^{1,2}$}
\author{Xiao Liang$^{1}$}
\author{Hui Zhai$^{1}$}

\affiliation{
$^1${Institute for Advanced Study, Tsinghua University, Beijing 100084, China}\\
$^2${Institute of Theoretical Physics and State Key Laboratory of Quantum Optics and Quantum Optics Devices, Shanxi University, Taiyuan 030006, China}}

\begin{abstract}
In this work, we propose a method for efficient learning of a multi-dimensional function. This method combines the Bayesian neural networks and the query-by-committee method. A committee made of deep Bayesian neural networks not only can provide uncertainty of the prediction but also can provide the discrepancy between committee members. Both the uncertainty and the discrepancy are large in the regions where the target function varies rapidly, and therefore, both quantities can be used to guide sampling data to such regions. In this way, we can learn a function accurately with the number of queried data points much less than uniform sampling. Here we test our method with two examples. One example is to find a rare phase in a phase diagram, which is separated from other phases by a second-order phase transition. In this example, the target function is the susceptibility function, and since the divergence of the susceptibility function locates the phase diagram, the task of searching such a phase perfectly matches the advantage of our method. Another example is to learn the distribution function for Monte Carlo integration of a high-dimensional function. In both examples, we show that our method performs significantly efficiently than uniform sampling. Our method can find broad applications in computational scientific problems.

\end{abstract}
\maketitle

\textit{Introduction.} In physics research, it is quite often that one encounters such kinds of problems of sampling a multi-dimensional space. For example, we are always interested in searching for exotic phases in a phase diagram spanned by multiple parameters of a Hamiltonian~\cite{Sondhi1997,Csontos2010,Chaikin1995}, and in many cases, the interested phase only occurs in a small parameter regime. Such examples include, for instance, the Fulde-Ferrell-Larkin-Ovchinnikov phase in a superconductor in the presence of the Zeeman field \cite{Casalbuoni2004,Matsuda2007,Kinnunen2018} and spin liquid phases in the frustrated magnets \cite{Norman2016, Zhou2017, Balents2017,Broholm2020}. For such situations, uniformly sampling parameter space can be quite low efficient. This problem is particularly serious when computing one point in the parameter space is already time-consuming. Another example is numerical integration over a multi-dimensional function, where the integrand function is usually highly peaked at one or several small regions or varies rapidly in certain small regions. In this situation, one needs to sample more points in these regions in order to obtain an accurate numerical integration. Various methods have been proposed to deal with such numerical integrations \cite{Davis1984,Kalos2008,Newman1999,Foulkes2001,Carlson2015}. The goal is always to obtain a numerical integration as accurately as possible, with minimal cost of computational resources. All these tasks essentially face the same issue, that is, how to sample multiple dimensional spaces efficiently.

There are also machine-learning-based methods to deal with this problem. One method is based on the Bayesian neural network (NN) \cite{Snoek2012,Shahriari2016,Wang2016}. Compared with the deterministic NN, a Bayesian NN not only predicts results but also provides the uncertainty of the prediction. The uncertainty is determined jointly by the prior probability of the parameters in NN and the likelihood of a NN with given parameters on the existing data. Usually in the region where the inferential uncertainty is large, it is hard for a NN to make conclusive predictions based on the existing data. Therefore, this inferential uncertainty can be used to guide adding more data points. It is usually efficient to add new data in the regions where the inferential uncertainty is large.

\begin{figure}[t]
    \centering
    \includegraphics[width=0.45\textwidth]{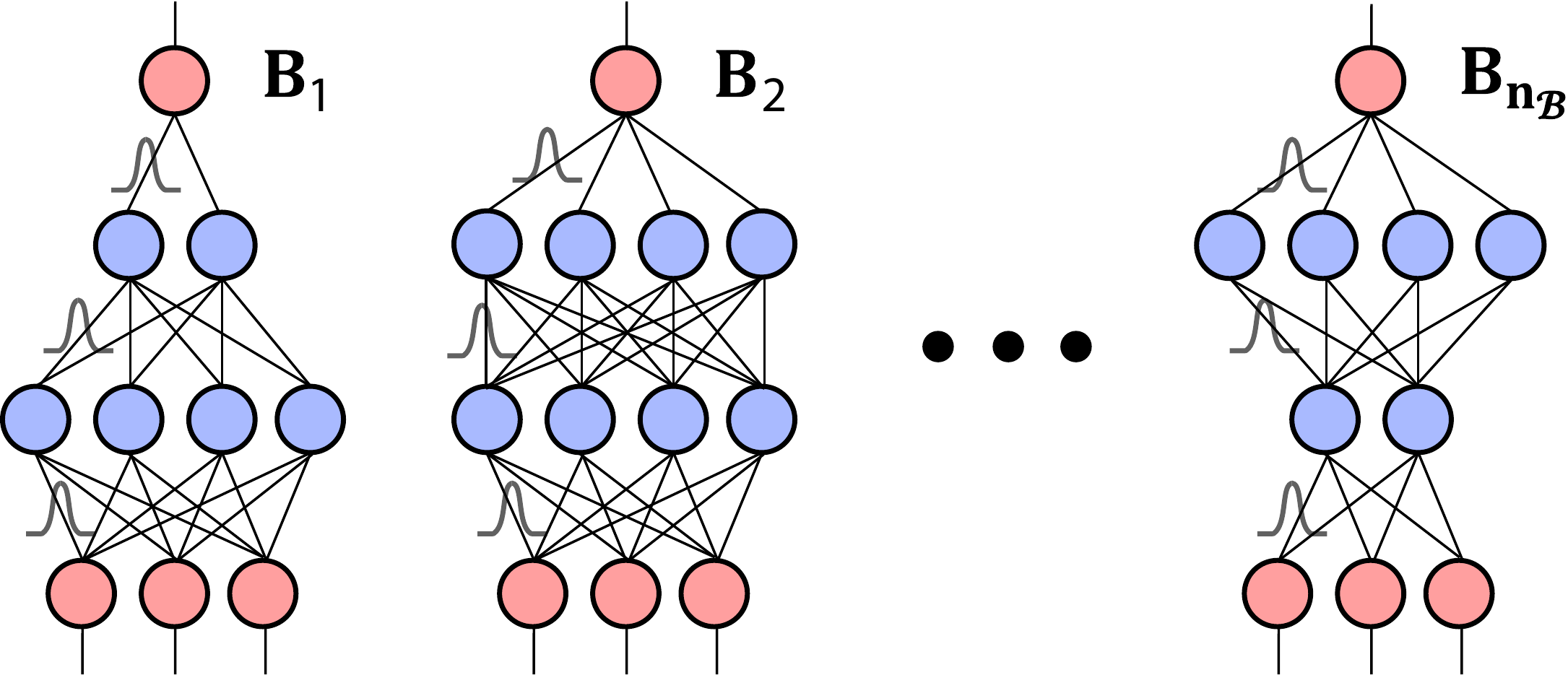}
    \caption{Schematic of a Committee made of the Bayesian NN. Different NN has different architectures, and the parameters in each BNN obeys a probability distribution, such that it not only gives predication, but also provides uncertainty of the predication. }
     \label{Committee}
\end{figure}

Another method is called query-by-committee method of active learning \cite{Abe1998,Settles2012,Seung1992,Freund1997,Zhao2006,Fu2013,Yao2020,Roy2001,Engelson1999}. A number of NNs with different architectures form a committee, and all these NNs are trained by the existing data. After training, all NNs make predictions on the entire parameter space, and the variance of predictions made by different NNs are computed. Usually in the region where the target function varies rapidly, it is hard for different NNs to reach a consensus and the variance is large there. Therefore, the variance can also be used to guide adding more data points. It is also efficient to add new data in the region where the variance is large.

In this work, to take advantage of both approaches, we propose a method that combines the Bayesian NN and the query-by-committee. We sample a multi-dimensional parameter space by iteratively adding data guided by both the inferential uncertainty of the BNN and the discrepancy among the committee members. We show two examples to demonstrate the advantage of this method in computational physics. One is searching for a rare phase in a phase diagram, and the other is a multi-dimensional numerical integration.

\textit{General Methods.} The task of our problem is to learn an $s$-dimensional function $y=f({\bf x})$, where ${\bf x}=\{x_1,x_2,\dots,x_s\}$ is an $s$-dimensional vector as input, and $y$ is a scalar as the output. We start with an initial dataset $\mathcal{D} = \{(\mathbf{x}_1,y_1),(\mathbf{x}_2,y_2),\cdots, (\mathbf{x}_{N_\mathcal{D}},y_{N_\mathcal{D}})\}$, where $N_\text{D}$ is the dataset size. Initially, $N_\text{D}=N_0$, and $N_0$ is a small number. At each round of learning, new data will be added and $N_\text{D}$ increases. The key of this method is to determine how to add the data efficiently. In other word, how to obtain a best fitting of the function with the smallest $N_\text{D}$.

As is shown in Fig. \ref{Committee}, we design $n_\mathcal{B}$ number of Bayesian NNs with different architectures and activations, labelled by $\text{BNN}_i$ $(i=1,\dots,n_\mathcal{B})$ (See appendix A for typical architectures of these Bayesian NNs). Each of the $\text{BNN}_i$ learns the existing data $\mathcal{D}$, and then, it returns a probability for output $y$ at certain input ${\bf x}$, which is denoted by $p_i(y|\mathbf{x},\mathcal{D})$. Then, the inference of each BNN is given by the mean value of $y$ weighted by the probability $p_i(y|\mathbf{x},\mathcal{D})$, i.e.
\begin{equation}
	\bar{y}_i(\mathbf{x}) = \int dy\left[y p_i(y|\mathbf{x},\mathcal{D})\right].
\label{Pi}
\end{equation}
In addition to the mean, $p_i(y|\mathbf{x},\mathcal{D})$ also carries the information of uncertainty of this inference. Obviously, for different NN, $p_i(y|\mathbf{x},\mathcal{D})$ are also different, and we can take an average over all $p_i(y|\mathbf{x},\mathcal{D})$, which yields
\begin{equation}
p_\mathcal{B}(y|\mathbf{x},\mathcal{D}) = \frac{1}{n_\mathcal{B}}\sum_{i=1}^{n_\mathcal{B}}p_i(y|\mathbf{x},\mathcal{D}).
	\label{pB}
\end{equation}

With the help of $p_i(y|\mathbf{x},\mathcal{D})$ and $p_\mathcal{B}(y|\mathbf{x},\mathcal{D})$, we can define the following two quantities. The first one is called a voting entropy $\mathbb{S}_\mathcal{B}$ defined by \cite{Settles2012,Zhao2006}
\begin{equation}
	\mathbb{S}_\mathcal{B}(\mathbf{x}) = -\int dy \left[p_\mathcal{B}(y|\mathbf{x},\mathcal{D})\ln p_\mathcal{B}(y|\mathbf{x},\mathcal{D}) \right],
	\label{Sb}
\end{equation}
and the second one is a relative entropy $\mathbb{K}_\mathcal{B}$ between individual inferential probabilities $p_i(y|\mathbf{x},\mathcal{D})$ and their average $p_\mathcal{B}(y|\mathbf{x},\mathcal{D})$, given by
\begin{equation}
	\mathbb{K}_\mathcal{B}(\mathbf{x}) = \frac{1}{n_\mathcal{B}}\sum_{i=1}^{n_\mathcal{B}}\int dy \left[ p_i(y|\mathbf{x},\mathcal{D})\ln\frac{p_i(y|\mathbf{x},\mathcal{D})}{p_\mathcal{B}(y|\mathbf{x},\mathcal{D})} \right].
	\label{KLB}
\end{equation}
$\mathbb{S}_\mathcal{B}$ and $\mathbb{K}_\mathcal{B}$ quantify different aspects of the prediction. The former accounts for committee averaged inferential uncertainties. The larger $\mathbb{S}_\mathcal{B}$, the larger the uncertainty is. The latter reflects the discrepancy of the inference made by different committee members. The larger $\mathbb{K}_\mathcal{B}$, the larger the discrepancy is. In terms of the Bayesian optimization, we should add new data in the region where $\mathbb{S}_\mathcal{B}$ is large. In terms of the query-by-committee, we should add new data in the region where $\mathbb{K}_\mathcal{B}$ is large. In our approach, we will simultaneously add new data in both regions.

\begin{algorithm}[t]
 \label{alg1}
 \caption{Query by the BNN Committee}

    \textbf{ Input:} Initial dataset $\mathcal{D}$ with $N_0$ initial data points \\
    {\textbf{ For} $n_r = 1,2,3,\cdots$
    \begin{enumerate}
    \item[\footnotesize{2.1}] Implement the Bayesian regression for each $\text{BNN}_i$ with the dataset $\mathcal{D}$; then output the inference $p_{i}(y|\mathbf{x},\mathcal{D})$ and $p_\mathcal{B}(y|\mathbf{x},\mathcal{D})$ \label{I1}
    \item[\footnotesize{2.2}] With $\mathbb{S}_\mathcal{B}(\mathbf{x})$ and $\mathbb{K}_\mathcal{B}(\mathbf{x})$, construct a set $\mathbf{X}_\text{add}$ which contains $N_\text{add}$ unlabeled $\mathbf{x}$ \label{I1}
    \item[\footnotesize{2.3}]  Query the labels for all $\mathbf{x} \in \mathbf{X}_\text{add}$, which form an adding dataset $\mathcal{D}_\text{add}$, and then add $\mathcal{D}_\text{add}$ into $\mathcal{D}$, i.e. $\mathcal{D} \rightarrow \mathcal{D} \cup  \mathcal{D}_\text{add}$ \label{I1}
    \item[\footnotesize{2.4}] \textbf{If} converge, \textbf{break}
    \end{enumerate}
    \textbf{ End for}}\\
    \textbf{ Output:} Final inferences $p_{\mathcal{B}}(y|\mathbf{x},\mathcal{D})$, $\bar{y}_\mathcal{B}$ and $\bar{y}^*_\mathcal{B}$.
\end{algorithm}

\textit{Pseduo-Code:} We explicitly present the pseudo-code of our method in the Algorithm~\ref{alg1}, where the round iteration begins from the \textbf{for} loop in the line 2. Here we present a bit more explanations on the Algorithm~\ref{alg1}.

\begin{itemize}
  \item 2.1: In this step, we train each $\text{BNN}_i$ and return the prediction by a method called the variational inference, which is an efficient method for the Bayesian inference with a deep NN containing a large number of neurons (See appendix A for details).
  \item 2.2: In this step, we select a set $\mathbf{X}_\text{add}$ of unlabeled points for label query. As discussed above, $\mathbf{X}_\text{add}$ is composed by three parts
\begin{equation}
\mathbf{X}_\text{add} = \mathbf{X}_\mathbb{S} \cup \mathbf{X}_\mathbb{K} \cup \mathbf{X}_\mathbb{R},
\label{add_data}
\end{equation}
where $\mathbf{X}_\mathbb{S}$ with size $N_\mathbb{S}$ contains points with large $\mathbb{S}_\mathcal{B}(\mathbf{x})$ and $\mathbf{X}_\mathbb{K}$ with size $N_\mathbb{K}$ contains points with large $\mathbb{K}_\mathcal{B}(\mathbf{x})$. $\mathbf{X}_\mathbb{R}$ with size $N_\mathbb{R}$ represents the points that are randomly sampled, which accounts for random exploration at each round of learning. Consequently, the adding point number in each round of learning is $N_\text{add} = N_\mathbb{S} + N_\mathbb{K} + N_\mathbb{R}$. Note that including $N_\mathbb{R}$ is necessary, which is to avoid the situation where most NNs are trapped in local minima.
  \item 2.3: In this step, we query the realistic labels $y$ for all points in $\mathbf{X}_\text{add}$, which forms an added dataset $\mathcal{D}_\text{add}$. Then, we add $\mathcal{D}_\text{add}$ into the total dataset.
  \item 2.4: In this step, we discuss the convergence condition, which serves as the stopping criterion for the round iteration. The basic idea is that the convergence is reached when the prediction does not change as new data points were added. In practice, we define $\bar{\mathbf{x}}$ as
\begin{equation}
\bar{\mathbf{x}} = \int d^s\mathbf{x} \left[\bar{y}^*_\mathcal{B}(\mathbf{x})\mathbf{x} \right],
\label{barx}
\end{equation}
where
\begin{align}
&\bar{y}_\mathcal{B}(\mathbf{x}) = \int dy\left[y p_\mathcal{B}(y|\mathbf{x},\mathcal{D})\right],\label{yB}\\
&\bar{y}^*_\mathcal{B}(\mathbf{x}) = \frac{|\bar{y}_\mathcal{B}(\mathbf{x})|}{\int d^s\mathbf{x} |\bar{y}_\mathcal{B}(\mathbf{x})|}.
\label{ystar}
\end{align}
and we find that $\bar{\mathbf{x}}$ works well as a convergence criterion. By definition, $\bar{\mathbf{x}}$ is simply the expectations of $\mathbf{x}$ weighted by $\bar{y}^*_\mathcal{B}(\mathbf{x})$. We stop the iteration in several rounds after $\bar{\mathbf{x}}$ converges.
\end{itemize}

\begin{figure}[t]
    \centering
    \includegraphics[width=0.48\textwidth]{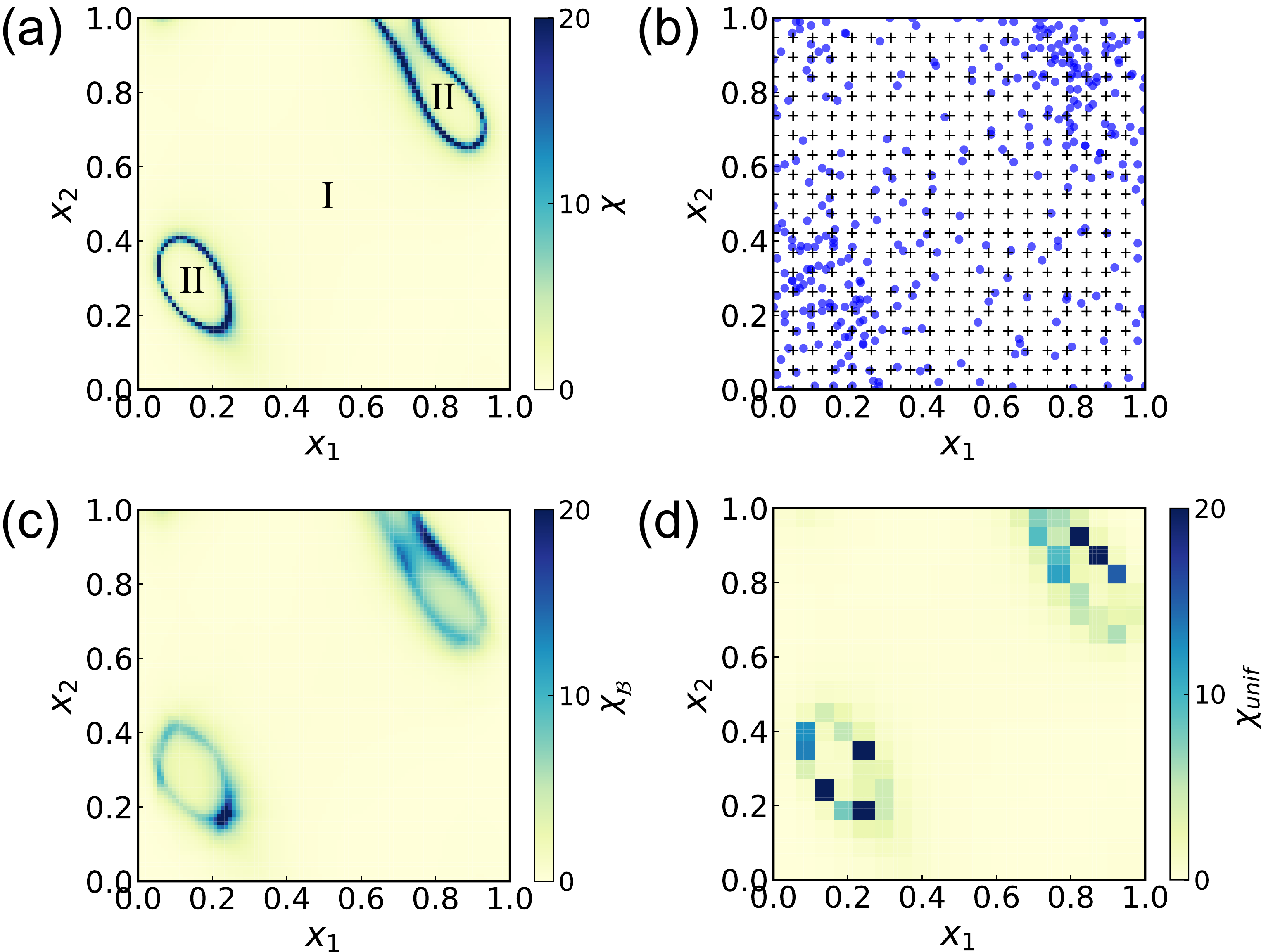}
    \caption{(a) The target phase boundary. (b) The data points sampled by the Bayesian Committee (blue dots) and the uniform sampling strategy (black plus signs). (c) The phase boundary obtained by the Bayesian Committee at $n_r = 30$ rounds of iterations. (d) The phase boundary obtained by uniformed sampling with the same number of data points. }
     \label{phase}
\end{figure}

\textit{Example I: Determining Phase Boundary.} Here we present the first example of searching a rare phase, which is separated from other phases by a second-order phase transition. Without loss of generality, we consider a $\phi^4$ model for the demonstration purpose, which describes a generic phase transition to a phase breaking a $Z_2$ symmetry \cite{Chaikin1995}. For the $\phi^4$ model, the free-energy in terms of the order parameter $\phi$ is given by
\begin{equation}
F = \frac{1}{2}\alpha \phi^2 + \beta \phi^4,
\label{free-energy}
\end{equation}
where $\beta$ is taken to be positive, and $\alpha$ is a function of model parameters ${\bf x}=\{x_1,x_2\}$. The transition from the $Z_2$-symmetry preserved phase I ($\alpha>0$ and $\phi=0$) to the $Z_2$-symmetry broken phase II ($\alpha<0$ and $\phi\neq0$) takes place at $\alpha=0$. Our goal is to search a small $Z_2$-symmetry broken phase II in a large phase space spanned by ${\bf x}$.

\begin{figure}[t]
    \centering
    \includegraphics[width=0.36\textwidth]{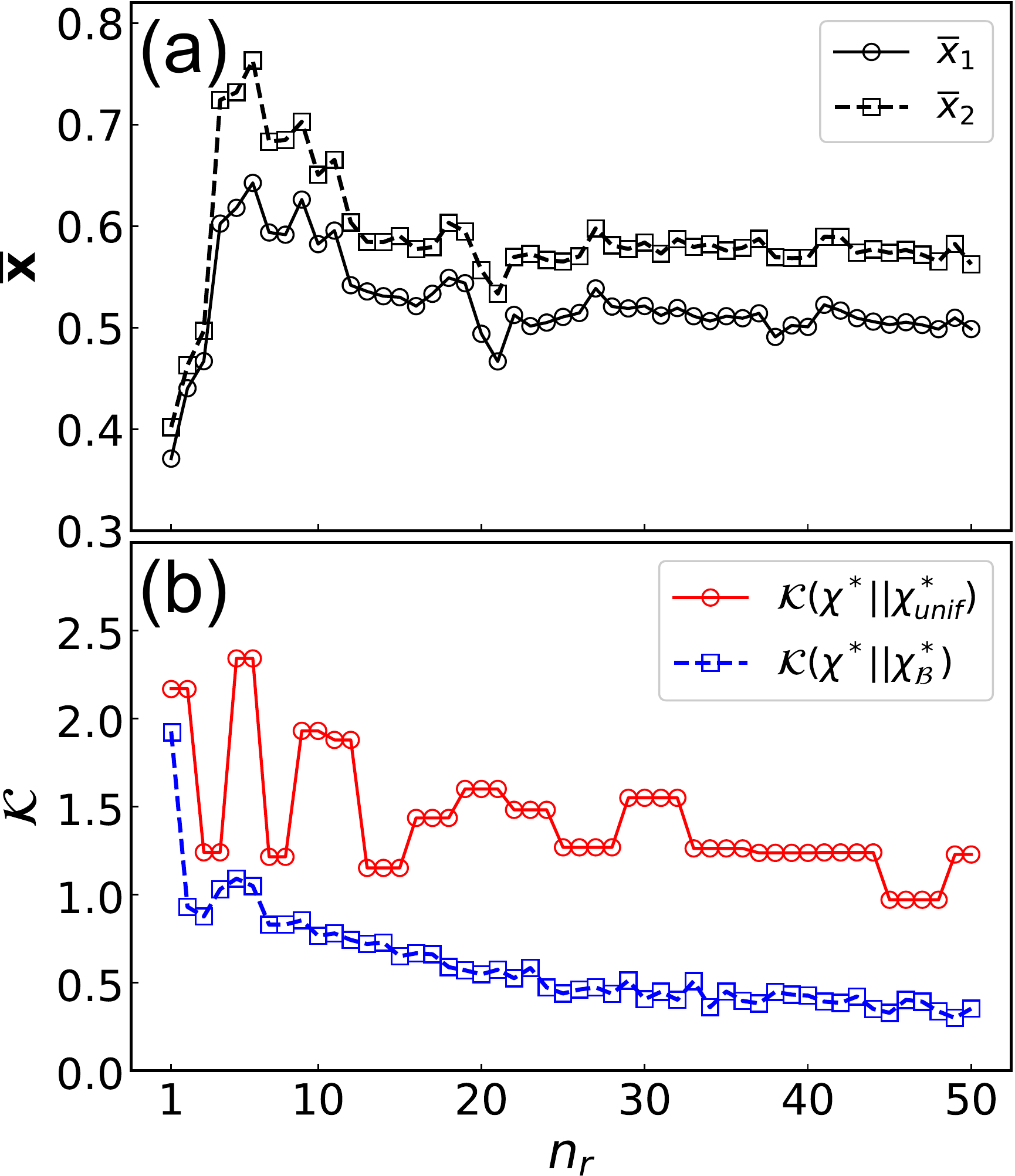}
    \caption{(a) $\bar{\mathbf{x}}$ as a function of the round number $n_r$. (b) The Kullback-Leibler divergences $\mathcal{K}(\chi^* || \chi^*_\mathcal{B})$ and $\mathcal{K}(\chi^* || \chi^*_\text{unif})$ as a function of the round number $n_r$. }
     \label{KL1}
\end{figure}

Conventionally, people determine the phase transition by directly looking at the behavior of the order parameter $\phi$. However, this is not effective for the learning process since $\phi$ is zero in large areas of the phase space. Therefore, instead of $\phi$, we consider the susceptibility $\chi$, which is given by \cite{Chaikin1995}
 \begin{equation}
\chi = \left\{
\begin{array}{rcl}
&\alpha^{-1}  \    \ \    \ \   \  \  \ \text{if }\alpha>0, \\
&(2|\alpha|)^{-1}  \   \  \text{if }\alpha<0.
\end{array}
\right.
\label{sus}
\end{equation}
We target to learn the function $\chi({\bf x})$, which diverges at the phase boundary. The divergence of the susceptibility $\chi$ indicates the tendency toward forming certain kinds of order. This behavior of susceptibility allows us to sense a second-order phase transition even far away from the phase boundary. The fact that $\chi$ is quite singular at the phase boundary making this problem perfectly suitable for our method since both the inferential uncertainty and the committee variance are large there. Hence, our method can naturally guide sampling more data into the region near the phase boundary, which helps to determine the phase boundary more efficiently.

In practices, the susceptibility can be calculated from the microscopic Hamiltonian \cite{Chaikin1995}. Here, to demonstrate our method generically, we choose $\alpha({\bf x})$ as a complicated function such that the phase boundary contains two disjointed pieces, as is shown in Fig.~\ref{phase}(a). (See Appendix B for the detailed form of $\alpha({\bf x})$) Then, we follow the Algorithm~\ref{alg1} described above by setting the size of the initial dataset as $N_0=50$. At each round of iteration, we add $10$ new data points with $(N_\mathbb{S},N_\mathbb{K},N_\mathbb{R})=(4,4,2)$. In Fig.~\ref{KL1}(a), we plot ${\bar x}$ defined in Eq.~(\ref{barx}) as a function of the round number $n_\text{r}$, in which we can see that ${\bar {\bf x}}$ saturates as $n_\text{r} \gtrsim 20$. Stopping at round $n_\text{r}= 30$, we show all the sampled data points in Fig.~\ref{phase}(b) by dots, where the total number of queried data points is $N_0+(N_\mathbb{S}+N_\mathbb{K}+N_\mathbb{R})n_\text{r} = 350$. As one can see that these data points are mostly concentrated in the region of the targeted phase II, which leads to the NN inferential phase boundary shown in Fig.~\ref{phase}(c). If we sample the same number of data points uniformly, the resulting phase boundary is shown in Fig.~\ref{phase}(d) whose quality is obviously lower than Fig.~\ref{phase}(c) obtained by our approach. To quantify the performance of our approach, we compute the Kullback-Leibler divergences between the real $\chi^*({\bf x})$ and the inferential $\chi_\mathcal{B}^*({\bf x})$ obtained by our method in Fig.~\ref{KL1}(b), i.e. $\mathcal{K}(\chi^* || \chi^*_\mathcal{B})$, where $\chi^*$ denotes the normalized susceptibility defined by Eq.~(\ref{ystar}). In comparison, we also plot the Kullback-Leibler divergence between the real $\chi^*({\bf x})$ and $\chi^*_\text{unif}({\bf x})$ obtained by uniform sampling, namely $\mathcal{K}(\chi^* || \chi^*_\text{unif})$. The smaller $\mathcal{K}$ is, the more similar the two functions are. One can clearly observe that, $\mathcal{K}(\chi^* || \chi^*_\mathcal{B})$ decreases much faster than $\mathcal{K}(\chi^* || \chi^*_\text{unif})$.

\textit{Example II: Numerical Integration.} Here we show our method can also be used for the Monte Carlo integration \cite{Kalos2008,Newman1999,Foulkes2001,Carlson2015}. Monte Carlo integration is to calculate an integral $I= \int d^s \mathbf{x} f(\mathbf{x})$ through sampling, and the key of such integration approach is to find a proper distribution function for sampling, which should work better than uniform sampling. Our strategy is to first learn $f(\mathbf{x})$ by the Bayesian committee, and then the normalized network inference $\bar{y}^*_\mathcal{B}(\mathbf{x})$ given by Eq.~(\ref{ystar}) can play the role of the sampling distribution function, i.e.
\begin{equation}
I_\mathcal{B} = \int d^s \mathbf{x} \bar{y}_\mathcal{B}^*(\mathbf{x}) \left[\frac{f(\mathbf{x})}{\bar{y}_\mathcal{B}^*(\mathbf{x})}\right].
\label{Ib}
\end{equation}

\begin{figure}[t]
    \centering
    \includegraphics[width=0.36\textwidth]{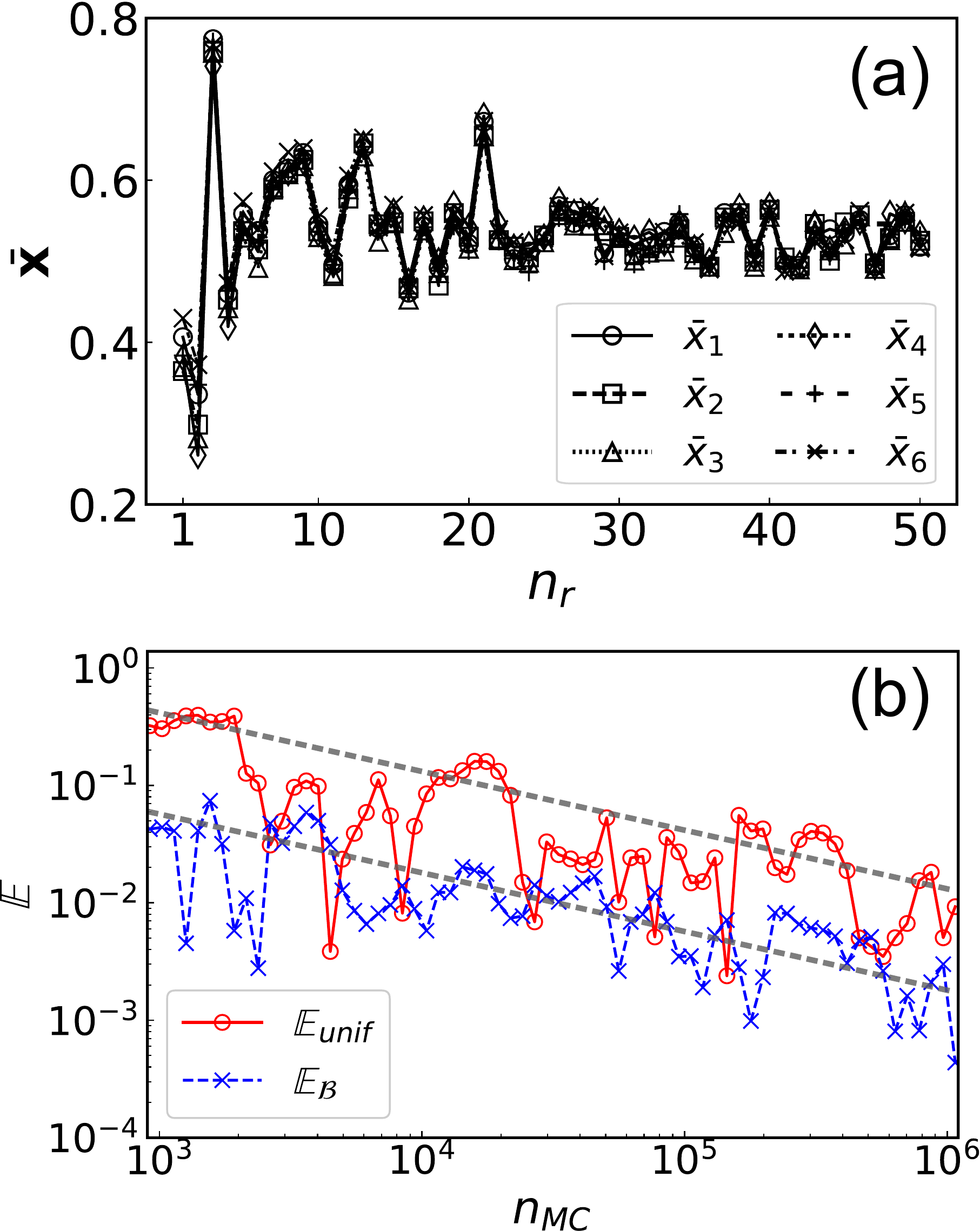}
    \caption{(a) The dependence of $\bar{\mathbf{x}}$ on the round number $n_r$.
    (b) Comparison of the Monte Carlo results obtained by our approach $\mathbb{E}_\mathcal{B}$ (dahsed line with crosses) and the fully random sampling $\mathbb{E}_\text{unif}$ (solid line with circles) as functions of the Monte Carlo sampling number $n_\text{MC}$.}
     \label{KL2}
\end{figure}

To explicitly demonstrate our method, we choose the integrand to be a complicated $6$-dimensional special function, which varies a lot in all dimensions and also possesses several sharp peaks. Note that, with the help of the properties of the special function, this integration can still be done analytically, which yields an exact value $I_\text{real}$ serving as a reference for our numerical results (see Appendix B for details). Following the procedurals of the Algorithm~\ref{alg1}, we take $N_0=10^4$, and $N_\mathbb{S} = N_\mathbb{K} = 2N_\mathbb{R} = 400$. We show our learning and Monte Carlo integration results in Fig.~\ref{KL2}. Fig.~\ref{KL2}(a) shows that $\bar{\mathbf{x}}$ becomes stabilized as $n_r\gtrsim30$. Then, we can stop at $n_\text{r}=35$, with totally $4.5\times 10^4$ points queried, and use the normalized inference at this round to guide Monte Carlo sampling. In Fig.~\ref{KL2}(b), we display the relative error of the Monte Carlo integration $\mathbb{E}_\mathcal{B}=\left|I_\mathcal{B}-I_\text{real}\right|/I_\text{real}$ obtained by our method as a function the Monte Carlo sampling number $n_\text{MC}$, and compare it with the relative error $\mathbb{E}_\text{unif}$ obtained by the uniform sampling Monte Carlo.  One can clearly see that, both errors $\mathbb{E}_\text{unif}$ and $\mathbb{E}_\mathcal{B}$ decease as $\gamma/\sqrt{n_\text{MC}}$ when $n_\text{MC}$ increases, which is expected for Monte Carlo integration. However, the coefficient $\gamma$ with uniform sampling is about $9$ times larger than $\gamma$ obtained by our method. That is to say, to reach the same accuracy, the required Monte Carlo sampling points with uniform sampling is roughly two orders of magnitude more compared with our method.

\textit{Summary and Outlook.} In summary, we have demonstrated a new method for efficiently sampling a high-dimensional function by combining the Bayesian NN and the query-by-committee. As a proof-of-principle demonstration, we have compared our method with uniform sampling in two examples discussed above, which have shown significant advantages. Here we emphasize that, when applying our method to real problems, there are always a number of things that can be further optimized. The complicity of NN structure can be adjusted based on the complicity of the problem and data structure, and the number of committee members can also be adjusted by the computational cost of learning. All the parameters such as $N_\mathbb{S}$, $N_\mathbb{K}$ and $N_\mathbb{R}$ can also be chosen properly depending on the balance between the computational cost of querying data and the size of the parameter space. With all these considerations, one can try to reach the most efficient sampling guided by our method, which can find broad applications in various kind of computational physics problems, as well as in computational tasks in other science problems.

\textit{Acknowledgement}. We thank helpful discussions with Wei Zheng, Yadong Wu, Juan Yao, Zhiyuan Yao and Ce Wang.  This work is supported by NSFC (Grant No. 11804205 (L.C.) and Grant No. 11734010 (H.Z.)), Beijing Distinguished Young Scientist Program, and MOST (Grant No. 2016YFA0301600).

\begin{appendix}

\section{Bayesian Neural Network: Structure and Learning}
In both examples shown in the main text, we adopt a Bayesian NN committee with $12$ Bayesian NNs \cite{MacKay1992,MacKay1995,Buntine1991,Blundell2015}. For each Bayesian NN, there are three hidden layers with the number of neurons in each layer randomly selected in the range of $2-20$. Furthermore, the activation function of each hidden layer is also randomly picked up between 'Relu' and 'tanh'. For the output layer, we choose the 'Relu' activation since the target functions are non-negative for both examples.

For each Bayesian NN, we establish a probabilistic model by introducing the following two parts of uncertainties \cite{Blundell2015}: First, the neural parameters $\Theta = \{\hat{w},\mathbf{b}\}$, with $\hat{w}$ and $\mathbf{b}$ corresponding to the weights and biases, satisfy a probability distribution $p(\Theta)$ called the prior distribution, which is simply taken as a joint normal distribution $p(\Theta) = \prod_{\theta\in\Theta} \mathcal{N}_\theta(0,1)$ in our calculation. Secondly, the output of the BNN also satisfies a modeling distribution $p(y|\Theta,\mathbf{x})$, which is commonly assumed to be in a Gaussian form, i.e.
\begin{equation}
p(y|\Theta,\mathbf{x}) = \frac{1}{\sqrt{2\pi}\sigma}\exp(-(y-y_f)/2\sigma^2),
\tag{A1}
\label{A1}
\end{equation}
where $y_f$ is the BNN's output of the final layer when $\Theta$ and $\mathbf{x}$ are given, and $\sigma$ is a hyper-parameter accounting for the intrinsic noise in the dataset $\mathcal{D}$. This kind of modeling naturally incorporates the neural networks into the framework of the Bayes's theorem. Note that, since $\mathcal{D}$ contains no intrinsic noise for both examples shown in the main text, we set $\sigma = 0.05$ to be a small value. Then, given a dataset $\mathcal{D}$, we can make the Bayesian inference
\begin{equation}
p(y|\mathbf{x},\mathcal{D}) = \int d\Theta \left[p(\Theta|\mathcal{D})p(y|\Theta,\mathbf{x})\right],
\tag{A2}
\label{A2}
\end{equation}
where $p(\Theta|\mathcal{D})$ is the posterior distribution that can be obtained by the Bayesian equation, i.e.
\begin{equation}
\begin{aligned}
p(\Theta|\mathcal{D}) &= \frac{p(\mathcal{D}|\Theta)p(\Theta)}{p(\mathcal{D})}, \\
&\propto p(\mathcal{D}|\Theta)p(\Theta), \\
&= \prod_{(\mathbf{x}',y') \in \mathcal{D}} p(y'|\Theta,\mathbf{x}') p(\Theta).
\end{aligned}
\label{A3}
\tag{A3}
\end{equation}
As one can see in Eq.~(\ref{A2}) that the Bayesian inference is usually computational challenging, especially for a deep BNN with large number of neurons, since one has to enumerate over the entire parametric space $\Theta$. This issue can be circumvented by the variational Bayesian inference \cite{Hinton1993, Graves2011,Blundell2015}. The variational inference is to first make a variational posterior distribution $q_\epsilon(\Theta)$ with $\epsilon$ being the variational parameters, and then minimize the Kullback-Leibler divergence
\begin{equation}
\mathbb{K}\left(q_\epsilon(\Theta)||p(\Theta|\mathcal{D})\right)  = \int d \Theta \left[ q_\epsilon(\Theta) \ln \frac{q_\epsilon(\Theta)}{p(\Theta|\mathcal{D})} \right].
\tag{A4}
\label{A4}
\end{equation}
When the minimization converges, the ansatz $q_\epsilon(\Theta)$ looks quite similar to the posterior $p(\Theta|\mathcal{D})$ such that one can simply replace $p(\Theta|\mathcal{D})$ by $q_\epsilon(\Theta)$ in Eq.~(\ref{A2}). Note that, as the ansatz $q_\epsilon(\Theta)$ is assumed to be some simple distributions, e.g. joint normal distribution, one can adopt the re-parameterisation technic \cite{Blundell2015,Opper2009,Kingma2014,Rezende2014} such that the minimization of Eq.~(\ref{A4}) can be done by the gradient descent optimization, since the re-parameterisation helps to keep the continuity of the computational graph. Practically in our calculation, we adopt this technic by assuming $q_\epsilon(\Theta)$ to be a joint normal distribution, i.e.
\begin{equation}
q_\epsilon(\Theta)  \propto \prod_{\theta\in\Theta}\exp(-(\theta-\mu_{\theta})/2\sigma_{\theta}^2),
\tag{A5}
\label{A5}
\end{equation}
with $\epsilon = \{\mu_{\theta},\sigma_{\theta}\}$ being the variational parameters.

\section{Calculation Details of the Two Examples}
\begin{figure}[t]
    \centering
    \includegraphics[width=0.48\textwidth]{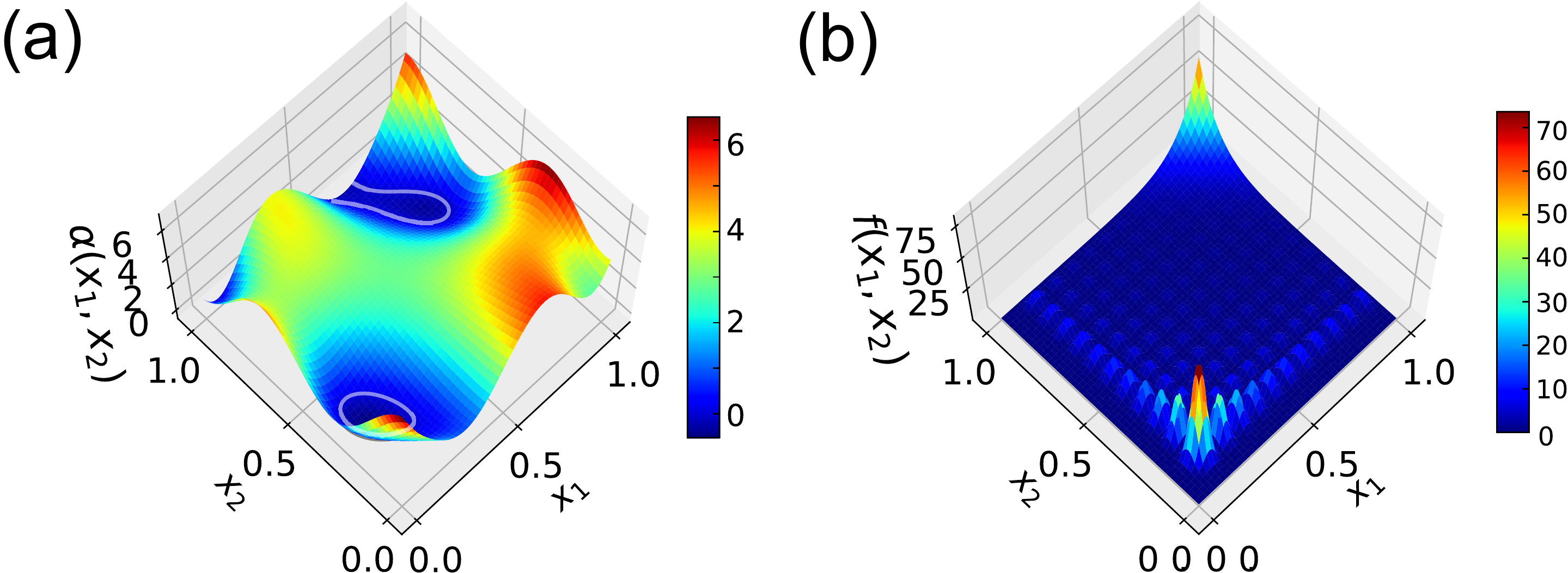}
    \caption{(a) $\alpha$ in the first example as a function $x_1$ and $x_2$. (b) Integrand function $f(x_1,x_2)$ in the second example as a function of $x_1$ and $x_2$, when $x_{3},\dots,x_6$ have been integrated out.}
     \label{FigA1}
\end{figure}
Here, we present more details for both examples. For the first example of learning the phase boundary, we generate the phase diagram Fig.~\ref{phase}(a) by a complicated $\alpha(x_1,x_2)$ as is shown in Fig.~\ref{FigA1}(a), where $\alpha=0$ indicates the phase boundary on which the susceptibility Eq.~(\ref{sus}) diverges. For the second example of Monte Carlo integration, the integrand is a complicated six-dimensional function generated by
 \begin{equation}
f(\mathbf{x}) = 10^{10}\left[\prod_{s=1}^6  \mathcal{J}^2_4(40 x_s) + \prod_{s=1}^6  e^{-10(1-x_s)}/5\right],
\tag{A6}
\label{A6}
\end{equation}
where $\mathcal{J}_4$ denotes the 4th-order Bessel function, and we choose the integration range within $x_s\in[0,1]$. Obviously, the integrand $f(\mathbf{x})$ is composed by two parts. The first Bessel function part oscillates quite fast in all dimensions with the major peak occurring at $x_s\approx 0.13$, and the second exponential function part is sharply peaked at $x_s = 1$. In Fig.~\ref{FigA1}(b), we show $f$ as a function of $x_1$ and $x_2$, as the rest $x_{s>2}$ have been integrated out. Due to the fact that integration over either the Bessel function or the exponent function can be done analytically, the integration over $f(\mathbf{x})$ can also be obtained analytically, which results in an exact integral value $I_\text{real} \approx  2.44461$, and this value serves as the reference of our numerical integration.

\end{appendix}

{}
\end{document}